# A Roadmap for Predictive Human Immunology

Aly A. Khan[1,2,3,4,*], Jason Perera[1], James Zou[5], Loïc A. Royer[6], Alan R. Lowe[7], Ambrose J. Carr[7], Theofanis Karaletsos[7], Patricia Brennan[7], Roham Parsa[8,9], Marcus R. Clark[10], Joe DeRisi[6], Jay Shendure[11,12,13,14], Sandra L. Schmid[6,15], Scott E. Fraser[16,17,18,19], Andrea Califano[8,20,21,22,23,24], Shana O. Kelley[1,25,26,27]

1. Chan Zuckerberg Biohub Chicago, Chicago, IL, USA
2. Departments of Pathology and Family Medicine, University of Chicago, Chicago, IL, USA
3. Toyota Technical Institute at Chicago, Chicago, IL, USA
4. Institute for Population and Precision Health, University of Chicago, Chicago, IL, USA
5. Department of Biomedical Data Science, Stanford University, Stanford, CA, USA
6. Chan Zuckerberg Biohub San Francisco, San Francisco, CA, USA
7. Chan Zuckerberg Initiative, Redwood City, CA, USA
8. Chan Zuckerberg Biohub New York, New York, NY, USA
9. Laboratory of Mucosal Immunology, The Rockefeller University, New York, NY, USA
10. Department of Medicine, University of Chicago, Chicago, IL, USA
11. Department of Genome Sciences, University of Washington, Seattle, WA, USA
12. Seattle Hub for Synthetic Biology, Seattle, WA, USA
13. Brotman Baty Institute for Precision Medicine, University of Washington, Seattle, WA, USA
14. Howard Hughes Medical Institute, Seattle, WA, USA
15. Department of Cell Biology, University of Texas Southwestern Medical Center, Dallas, TX, USA
16. Chan Zuckerberg Imaging Institute, Redwood City, CA, USA
17. Department of Biomedical Engineering, University of Southern California, Los Angeles, CA, USA
18. Translational Imaging Center, University of Southern California, Los Angeles, CA, USA
19. Molecular and Computational Biology Department, University of Southern California, Los Angeles, CA, USA
20. Department of Systems Biology, Columbia University, New York, NY, USA
21. Department of Biochemistry and Molecular Biophysics, Vagelos College of Physicians and Surgeons, Columbia University, New York, NY, USA
22. Department of Biomedical Informatics, Vagelos College of Physicians and Surgeons, Columbia University, New York, NY, USA
23. Herbert Irving Comprehensive Cancer Center, Columbia University, New York, NY, USA
24. Department of Medicine, Vagelos College of Physicians and Surgeons, Columbia University, New York, NY, USA
25. Department of Chemistry, Northwestern University, Evanston, IL, USA
26. Department of Biomedical Engineering, Northwestern University, Evanston, IL, USA
27. Robert H. Lurie Comprehensive Cancer Center, Northwestern University, Evanston, IL, USA
* Corresponding author

## Abstract

For over a century, immunology has masterfully discovered and dissected the components of our immune system, yet its collective behavior remains fundamentally unpredictable. In this perspective, we argue that building on the learnings of reductionist biology and systems immunology, the field is poised for a third revolution. This new era will be driven by the convergence of purpose-built, large-scale causal experiments and predictive, generalizable AI models. Here, we propose the Predictive Immunology Loop as the unifying engine to harness this convergence. This closed loop iteratively uses AI to design maximally informative experiments and, in turn, leverages the resulting data to improve dynamic, in silico models of the human immune system across biological scales, culminating in a Virtual Immune System. This engine provides a natural roadmap for addressing immunology's grand challenges, from decoding molecular recognition to engineering tissue ecosystems. It also offers a framework to transform immunology from a descriptive discipline into one capable of forecasting and, ultimately, engineering human health.

Creating an accurate and predictive model of the human immune system has long served as a guiding ambition for immunology. From early theoretical models[1-4] to the concept of "virtual cells"[5,6], the goal has always been to build a bridge from observation to prediction and, ultimately, to intervention. However, the realization of this goal remains elusive, fundamentally constrained by the complexity of countless cell types working in concert across diverse tissues and a focus on data and models that often capture what the system has done while failing to predict what it will do next. How might we overcome these constraints and chart a path forward?

Two great revolutions in immunology over the past century have brought us to our current understanding. The first, driven by the framework of reductionist biology, defined the elegance of the immune response through the discoveries of clonal selection[4] and somatic recombination that generate receptor diversity[7-9]; of MHC restriction and antigen presentation[10-12]; of the separation of B and T lymphocyte lineages[13] and their antigen receptors[14-16]; of the critical role of immune checkpoints[17-21] and regulatory T cells in maintaining peripheral tolerance[22-25]; and of dendritic cells[26,27], cytokines[28], and pattern-recognition receptors that link innate and adaptive immunity[29-31]. However, this reductionist approach, while foundational, struggled to predict the collective response of the immune system, motivating the need for a systems-level view.

The second revolution, born from the genomics era[32,33], ushered in systems immunology[34-38]. Powered by high-throughput single-cell technologies[39-47], we now have comprehensive "atlases" of cells and molecules across diverse tissues. These atlases have been instrumental in standardizing cell-type and state taxonomies[48-51], revealing rare or transient cell populations[52,53], elucidating tissue-resident immune programs and microenvironmental niches[54,55]. Despite these considerable achievements, most atlases are descriptive or correlative by design. As high-dimensional, static snapshots of a dynamic process, they make it challenging to derive causal insights or predict future states[56,57]. A transcriptomic signature that correlates with a cancer patient's response to anti-PD-1 therapy, for instance, could easily fail to generalize across cohorts; most importantly, it fails to provide mechanistic insights that could convert a non-responder into a responder[58,59] or forecast immune-related adverse events. Static maps are ill-suited to revealing the governing principles of the immune system's dynamics, or the bi-directional communication between cells, both of which are central to the immune response in vivo. This is the natural consequence of studying complex systems with tools optimized for either deep, low-throughput causal inference or shallow, high-throughput correlational analysis.

We are now poised for a third revolution, one that combines the power of artificial intelligence with rich data from large-scale experimental biology pipelines. In contrast to the previous two, this revolution will be driven not by a single technology, but by the convergence of two maturing fields. In biology, a new generation of purpose-built technologies can generate large-scale causal and perturbational data[60-65], moving us beyond correlational snapshots. In artificial intelligence, a profound shift toward causal and generative models offers unprecedented capabilities[66-68]. The success of AlphaFold[69,70] provides the crucial proof-of-concept, which was the result of deliberately aligning the right data with the right AI models to learn the evolutionary constraints of protein folding. This leap suggests that the grammar of immune recognition and the logic of cellular communication may now be solvable, provided the models are infused with the right biological knowledge and foundational data. Crucially, the success of this new era is not guaranteed unless we deliberately and iteratively connect data generation with model development in a principled way

Here, we propose the Predictive Immunology Loop as the unifying framework for this new era (Fig. 1). Our approach is built on two core principles: first, the deliberate alignment of fundamental immunological challenges, causal experimental paradigms, and advanced AI models (Fig. 1a); and second, an iterative, closed-loop engine where models actively generate non-obvious, testable hypotheses and guide subsequent experiments to maximize insight (Fig. 1b). This loop, from measurement to modeling to model-guided redesign, provides a roadmap for transforming immunology into an integrated engine for generating mechanistic and predictive understanding. This intentional framework will allow us to create a Virtual Immune System, moving from describing a system's states to predicting its trajectories and, eventually, to engineering its outcomes.

**The Data Imperative: Generating Causal Data to Power AI**

The predictive power of any AI model is fundamentally constrained by the data from which it learns. To build models that can infer the rules of immunity, we must generate large scale datasets that are purpose-built for causal inference. This requires a coordinated strategy that leverages the strengths of both human and animal model systems. While the ultimate ground truth lies in the human patient, a complete picture of the immune system can only be built by integrating three complementary sources of data. First, direct measurements from patients and healthy volunteers provide the essential clinical context. Second, because many perturbations are intractable in humans, model organisms remain indispensable platforms for longitudinal imaging and in vivo validation. Third, the gap between these two is now being bridged by sophisticated human model systems. Patient-derived organoids[71,72] and ex vivo tissues[73,74], for instance, allow for causal interrogation in a controlled setting. By integrating data from these sources, we can build a more complete and causally grounded picture of the human immune system. Ultimately, this integrated approach must measure the immune system along five critical axes where our functional understanding remains incomplete: genetics, molecular interactions, cellular decision making, tissue level organization, and system dynamics (Fig. 2).

**Genetics.** First, at the level of genetics, while genome-wide association studies (GWAS) have linked hundreds of loci to immune-mediated diseases, the gap from statistical association to mechanism remains vast[75,76]. Bridging such gaps requires purpose-built, causal datasets. For example, base editing can introduce disease-associated variants into primary human T cells[77-80] and pooled CRISPR screens can systematically perturb every regulatory factor[61-63,65,81]. Using single-cell multi-omics to read out the effects, such approaches can directly map how a genetic variant alters a cell's regulatory network. Datasets with this level of dimensionality provide ideal input for Transformer-based models to learn the complex, non-linear "grammar" of the non-coding genome, deciphering how distal enhancers regulate their target promoters[66,82]. More focused explorations will be needed to interrogate the genetic link of autoimmune diseases to the HLA locus and the antigen processing and presentation pathway. For example, immunopeptidomic comparison of CRISPR-engineered variants at these loci can reveal how genetic context shapes HLA-presented repertoires. Here, the predictive loop begins when a trained model, for example, hypothesizes that a specific variant alters a key transcription factor binding site. Subsequent base editing to engineer that variant into primary cells and experimentally measuring its functional consequence, such as T cell exhaustion, will provide a direct causal link that refines the model.

**Molecular.** Second, for molecular interactions, decades of structural biology have yielded exquisite views of how antibodies and T-cell receptors recognize their targets[12,83-85], but our ability to predict these interactions from sequence alone remains a frontier. The challenge here is one of data scale. Solving it requires a shift from sparse, individual structure solutions to dense, landscape-scale affinity mapping. High-throughput protein engineering platforms, such as yeast and phage display[86-88] coupled with deep mutational scanning and immunopeptidomics[89-92], and the advent of high-throughput cryo-EM now make this possible. For the AI community, this transforms a sparse classification problem into a dense regression task that maps sequence to a continuous landscape of interaction measurements[87]. This creates the ideal training substrate for Protein Language Models[93], which can learn the universal grammar of immune recognition and enable the rational design of novel receptors. The same principles apply to modeling cytokine receptors, adhesion molecules, and transcription factors. The loop here becomes a powerful engine for rational design. A model trained on this data could, for example, design a de novo TCR sequence predicted to bind a specific tumor neoantigen with high affinity. This prediction is tested by synthesizing the TCR and validating its function experimentally, with the results directly informing the next, more ambitious cycle of design.

**Cellular.** Third, to understand cellular decision making, we must capture the full trajectory of a cell's fate, as a single cell RNA seq endpoint reveals little about this dynamic journey. To model cell fate, we can use in vivo lineage tracing in model organisms[94] and deploy multiplexed perturbations with multi omic readouts in human cell systems[60-62,95,96]. These high dimensional time series and perturbative data are well suited for conditional variational autoencoders, generative adversarial networks or diffusion models. The power of these models lies in their ability to not only forecast a cell's fate but also to run "backwards in time" to computationally design the optimal sequence of signals required to steer a cell toward a desired state[67,97-104]. For example, a model might predict a precise temporal sequence of cytokine signals to drive T cell differentiation. These predictions can be tested with T cells in vitro, validating their functional state and moving immunology from observation to active control.

**Tissue.** Fourth, immune function is an emergent property of tissue ecosystems, where the spatial organization of cells and local interactions govern collective behavior. Restoring the spatial context lost in dissociated assays is therefore essential[105]. While most current spatial approaches are limited to static snapshots, a key frontier is the development of methods for longitudinal imaging and analysis to capture tissue dynamics over time. The imperative is to engineer immune-competent organoids or use tissue explants, such as from lymph nodes or tonsils, and to instrument them for longitudinal analysis with multiplex imaging and spatial omics[106]. High content imaging tools can provide a natural link between the states of a cell and its neighbors, tracking their lineages and their interactions over time[107,108]. Datasets that link each cell's molecular state and clonal lineage to its exact coordinates are naturally represented as graphs. Such data provides the ideal inductive bias for graph neural networks (GNNs)[109,110] to learn how a cell's state is influenced by its local niche, such as cell-to-cell proximity and chemokine gradients. A predictive GNN model of a germinal center, for example, can reveal the intercellular programs that shift outcomes toward high-affinity plasma and memory B cells. These in silico predictions are then tested by systematically perturbing these interactions in an ex vivo tissue explant, validating a model guided strategy for accelerating a vaccine response.

**Individual.** Finally, these scales converge at the level of the integrated individual. Here, the goal is to learn a personal "immunological set point"[111], from both host and environmental features. This requires a paradigm shift toward capturing continuous, multi-modal data streams. The AI challenge here is heterogeneous data fusion: integrating sparse, deep molecular measurements (e.g., periodic immune profiling)[112-114] with dense, continuous physiological data (e.g., from wearables) and clinical health records[115]. Multi modal AI models are designed for this specific task. Learning this personalized baseline allows us to reconceptualize disease as a quantifiable deviation. The loop is closed at the clinical level: the model forecasts a patient-specific disease trajectory or response to therapy. This prediction guides clinical intervention, and the patient's outcome provides the ultimate ground truth data to refine the model for future predictions. The overarching ambition is to create AI architectures that not only model each biological scale but also unify them into a single, predictive conceptual model of the immune system.

**Modeling to Infer the Rules of the System**

The causal data generated as described provides the necessary substrate, but the modeling stage will translate this data into understanding. The goal is a profound departure from the descriptive, clustering-based approaches that have dominated biological analysis. Instead of asking, 'what groups do these data fall into?', we must ask, 'what are the generative rules that could have produced these data?' This requires a new class of AI architectures, selected to match the intrinsic structure of the biological problem at hand.

Being precise about the model class required reveals a critical distinction. Much of biomedical AI has emphasized predictive modeling that fits correlations to forecast outcomes; such models are not, on their own, suitable for counterfactuals nor are they mechanistic. Causal models move beyond correlation to estimate the effects of interventions and support "what-if" reasoning. Generative models propose new candidates or actions; when constrained by causal structure and biological knowledge, they enable rational design. In the Predictive Immunology Loop, we ground powerful generative models in causal principles, moving beyond simple prediction toward design. Ultimately, the insights gained from these models will serve as the foundation for a unified, conceptual understanding of the immune system.

A foundational principle is that these models should be biologically informed. Rather than learning from a blank slate, the most successful architectures will capture an inductive bias that reflects a fundamental biological process. A critical component of this strategy is to ensure mechanistic interpretability, which requires developing methods to move beyond 'black box' predictions and help infer the novel biological principles the models have learned. Furthermore, embedding domain-specific priors, such as the known logic of signaling pathways or the graph-like structure of cell-cell communication, can be directly incorporated into the model's design. This approach ensures that generated outcomes are not only predictive but also realistic and grounded in validated, known biological knowledge. However, a critical balance must be struck between incorporating validated priors and allowing for unbiased discovery. Over indexing on prior knowledge carries the inherent risk of confirmation bias, as it may constrain the model from discovering truly novel biology not captured in existing knowledge.

For the **genetic** and **molecular** scales, the primary challenge is to learn the language of biological sequence. For this, Transformer architectures and the Protein Language Models built upon them provide a powerful approach. Their attention mechanism is ideal for representing the non-linear, long-range interactions that govern function, such as enhancer-promoter contacts in the genome or the folding of distal amino acids in a receptor. As these models are applied to immunology, predictive accuracy alone is insufficient; the central challenge is ensuring causal grounding. A model predicting a TCR-pMHC interaction, for example, must base its inference on relevant biophysical determinants rather than spurious correlates. Accordingly, we must integrate causal evaluation, e.g., tests of invariance across different TCRs/peptides, prospective mutational validation, and checks for mechanistic feature attribution, so that learned features correspond to verifiable mechanisms.

At the **cellular** scale, the goal is to model the dynamic landscape of decision-making and predict a cell's fate. Here, Conditional Diffusion Models offer a promising solution. Analogous to learning the entire Waddington landscape, these models capture the full probability distribution of cellular trajectories. Their power is unlocked through conditioning, where the process is steered by biologically meaningful inputs like cytokine exposure. This enables powerful counterfactual queries: not only what a cell does, but what it would do under a defined intervention. A key research frontier is to couple these models with Reinforcement Learning to learn optimal policies, such as an ordered sequence of stimuli, that guide a cell toward a desired therapeutic endpoint. Furthermore, because the function space of these models is so vast, preventing biologically implausible outputs remains a central challenge, necessitating the embedding of domain-specific priors to constrain the system.

At the **tissue** and **individual** scales, the challenge is to model the relational networks that govern emergent behavior. At the tissue level, Graph Neural Networks (GNNs) provide a natural architecture for this domain, where relationships between entities are as important as the entities themselves. In this framework, cells are represented as nodes whose states evolve through message passing with their neighbors, a direct computational analogue of cell-cell communication. This inductive bias is a perfect match for the structure of spatial omics data, preserving the neighborhood context lost in dissociated datasets. The true goal, however, is to embed these learned interaction rules into dynamic simulations, leading to a hybrid agentic framework where a GNN acts as the decision-making engine for each cell. The critical test for these frameworks is out-of-distribution generalization: they must produce robust predictions when exposed to novel perturbations, such as new checkpoint inhibitors. Success here would mark the transition to a true, executable model of the immune system, capable of predicting how an intervention reshapes an entire system.

**Predict, Hypothesize, and Engineer: Closing the Loop**

The transition from models back to experiments is the critical stage that closes the loop. The models we develop are not final outputs; they are dynamic, in silico laboratories for exploring the causal structure of the immune system. This process is operationalized through a forward pass of hypothesis generation and a dual-feedback pass of refinement.

The loop begins with a forward pass from model to experiment. Here, hypothesis generation is transformed from a human-driven process to a model-driven one, where models function as a virtual immune system to generate non-obvious, testable hypotheses. This moves beyond simple prediction to complex engineering. For instance, a GNN trained on cellular and molecular profiling data from an inflamed psoriatic skin biopsy can be subjected to in silico perturbations to identify the optimal ligand-receptor pair to disrupt a pathogenic feedback loop. More ambitiously, generative models can solve inverse problems, such as designing a novel TCR sequence with high affinity for a specific tumor antigen or predicting the precise signals required to steer an autoreactive T cell toward a stable regulatory phenotype. These computationally designed therapies are then validated at the bench, with the experimental outcome providing the essential error signal.

This error signal, which represents the mismatch between prediction and reality, drives a dual feedback mechanism for learning and refinement (Fig. 1b). It generates a response to activate an inner feedback loop focused on computational and analytical improvement. Here, we interrogate the model and data to determine if the failure can be remedied with existing tools and knowledge. We ask if the model architecture is suboptimal, if the training data is incomplete, or if we are missing a critical biological prior. For example, if a designed TCR shows unexpected off-target toxicity, this inner loop would guide the retraining of the model with more comprehensive data or the integration of new biophysical constraints to better penalize cross-reactivity. This process iteratively refines our in silico representation of biological reality.

Persistent failures that cannot be resolved through this inner loop suggest a more profound problem: a gap in our ability to observe the system. This activates a powerful outer feedback loop, which transforms a model's failure into a mandate for technological innovation. The feedback loop asks if the prediction error points to a need for a completely new kind of measurement. If models consistently fail to predict the long-term dynamics of cellular states, it may indicate that static, single-point assays are fundamentally insufficient. This failure would then directly inspire the development of new platforms capable of longitudinal, multi-omic tracking of single cells in situ. The proposed dual-loop process uses errors to first refine our models and then to invent new measurement technologies. Crucially, an exciting promise of AI is its potential to help us design new experiments and technologies, not just make sense of existing data. This creates a complete, self-improving system that propels both our understanding and our ability to engineer human health.

**The Path Forward: Building an Ecosystem for Predictive Immunology**

The predictive immunology loop will feel familiar to systems biologists, but its coordinated, multi-scale application, with causal measurements and design-oriented AI at each step, is the novelty we advance here. Realizing the vision of the Predictive Immunology Loop requires more than just technological maturation; it demands a radical change in the culture of how we conduct our science, train our students, and collaborate across disciplines. Some of the longest standing grand challenges of immunology, deciphering the rules of inflammation, designing a universal influenza vaccine, curing autoimmunity, are systems-level problems that exceed the capacity of any single laboratory or even a single discipline.

The foundation of this ecosystem will be a new generation of scientists. The future of immunology will be driven by researchers who are multilingual, possessing deep, mechanistic intuition in immunology while being computationally fluent. This requires a deliberate rethinking of our graduate and postdoctoral training paradigms. While existing fellowship and early-career programs have begun to bridge these disciplines, we must now establish truly integrated training programs where students learn to co-design experiments and computational models from day one, rotating through both wet and dry labs as a core part of their education. The goal is not to create experimentalists who can run a pre-packaged software tool, but to cultivate a generation of scientists who can reason from first principles in both domains.

Second, this new science requires a new model of collaboration. The scale and complexity of generating the necessary training and benchmark datasets and validating the resulting models call for large, multi-investigator consortia. This is a call to action for funding organizations, which are uniquely positioned to catalyze progress. Their support will be essential in establishing shared, open-source infrastructure, including repositories of validated models and the benchmark datasets essential for their training. Creating these foundational resources is critical to fuel the rapid advancement of new methods, encourage the active participation of the machine learning community, and democratize access, allowing the broader scientific community to build upon a common, rigorously validated foundation.

Third, to focus collaboration and benchmark progress, we propose a set of 'Immune Turing Tests': pre-registered, prospective challenges designed to push the boundaries of predictive and generative modeling. Inspired by the successes in protein structure prediction, these tests would establish concrete goals with transparent metrics. Challenges would span the scales of human immunology, from predicting the functional consequences of genetic variants of unknown significance in allergy, autoimmunity, and immunodeficiency, to inverse-designing TCRs and antibodies with specified affinity for novel antigens de novo, and to developing tissue-level "world models" that forecast spatial reorganization and collective function under defined perturbations, such as in a germinal center. Each benchmark must be adjudicated against held-out biology using transparent metrics, pairing forward prediction with design-in-the-loop validation.

Finally, the private sector has a vital role in translating these insights into impact. The "Engineer" stage of the loop aligns directly with the goals of the biotechnology and pharmaceutical industries, and public-private partnerships will be crucial for developing the rationally designed immunotherapies this framework enables. A century of immunology research has laid an incredible foundation. By embracing this iterative and collaborative approach, we can move from understanding the immune system to engineering its outcomes for human health.

Beyond the immediate goals of prediction and engineering, this new era offers the prospect of a unified conceptual model of the immune system, one that provides understanding with the same elegance as the Standard Model in Physics or the central dogma of molecular biology. The ultimate test of the framework we propose will be its ability to synthesize the vast and disparate knowledge of immunology into such a coherent set of principles. The emerging capacity of artificial intelligence to reason over immense bodies of scientific knowledge and orchestrate the interactions between individual, scale-specific models suggests this grand challenge may now be within reach, representing a fundamental leap in our ability to move from understanding the immune system to engineering its outcomes for human health.

**Figure 1**

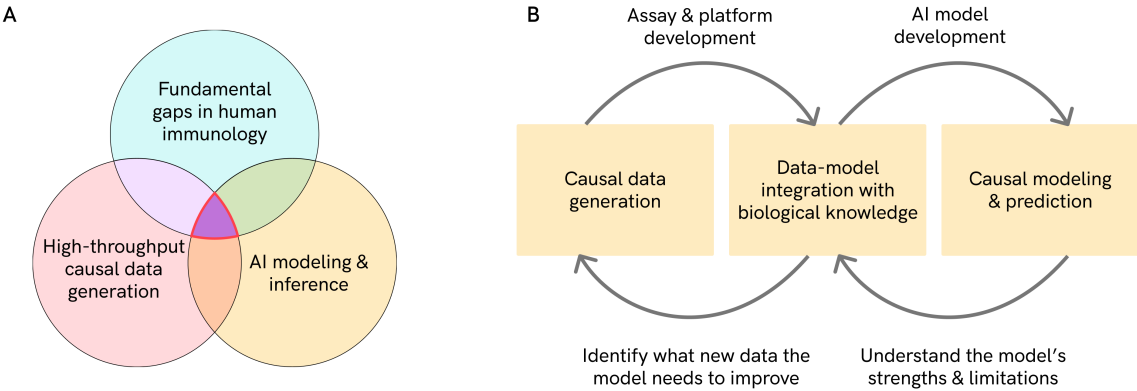

Figure 1 – The Predictive Immunology Loop: A Framework for Discovery and Design

**Figure 1: The Predictive Immunology Loop: A Framework for Discovery and Design.**

(A) A new frontier in immunology is opened by the convergence of three critical domains: defining fundamental gaps in immunological knowledge, generating large-scale causal data, and deploying predictive AI models. Their intersection creates a powerful engine for discovery and hypothesis generation. (B) This engine is operationalized as an iterative, closed loop. Causal data from advanced assays fuel the development of biologically-informed AI models, which in turn generate causal predictions. The loop is closed by a dual feedback mechanism: model performance analysis guides iterative refinement, while identifying critical data gaps directs the next wave of experimentation. This framework is designed to accelerate the transition from observation to rational intervention in human health.

**Figure 2**

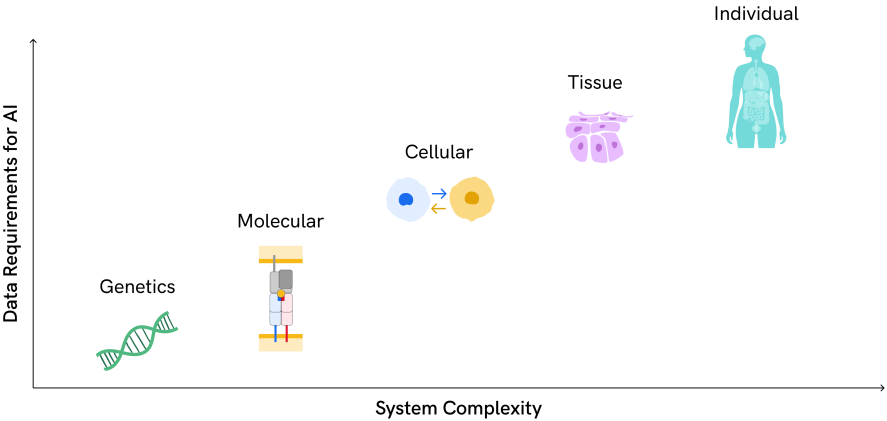

Figure 2 - Modeling the Full Continuum of Immunity

**Figure 2: A Multi-Scale Roadmap for Predictive Immunology.**

Modeling the human immune system requires integrating data across vastly different biological scales, from the genetic blueprint to the integrated individual. As the system's complexity increases, so too do the demands on data dimensionality and structure required for building predictive AI models. This roadmap highlights the critical need for purpose-built, causal datasets at each scale, from molecular recognition and cellular dynamics to tissue ecosystems, to train powerful AI architectures capable of capturing the emergent rules of immunity and ultimately forecasting patient-specific trajectories.

**Table 1**

| Scale of Inquiry | Genetic |
|---|---|
| **Core Question** | What is the mechanistic basis of known risk alleles in the coding and non-coding genome that drive immune diseases and deficiencies? |
| **AI-Enabled Problem** | To learn the gene-to-function grammar that links genetic variants to the regulation of immune genes in specific cell types and states. |
| **Key Data Inputs** | Population-scale GWAS; single-cell multi-omics (scRNA/ATAC-seq) from patient cohorts; high-throughput in vivo/organoids, CRISPR screens to validate causal and functional links. |
| **Representative AI Models** | Transformers that excel at learning the long-range dependencies in the "language" of the non-coding genome, mirroring how distal enhancers regulate gene promoters. |
| **Therapeutic Opportunity** | Mechanistic Basis of Genetic Risk: Moving from statistical correlation to a mechanistic understanding of disease drivers, enabling novel drug targets and patient stratification based on precise disease-causing pathways. |
| **Scale of Inquiry** | Molecular |
| **Core Question** | How does protein sequence encode the specificity and affinity required for receptor-antigen interactions? |
| **AI-Enabled Problem** | To predict the binding affinity and specificity of any TCR-antigen or antibody-antigen pair directly from their primary amino acid sequences. |
| **Key Data Inputs** | High-throughput affinity mapping (yeast/phage/mammalian cell display); deep mutational scanning; large-scale structural data (Cryo-EM) of immune receptor complexes. |
| **Representative AI Models** | Protein Language Models (PLMs): Pre-trained on all known proteins, these models learn the fundamentals of protein folding, allowing fine-tuning to predict the specific rules of immune recognition. |
| **Therapeutic Opportunity** | Molecular Recognition & Rational Design: Designing high-affinity TCRs that precisely target patient-specific neoantigens, and creating novel cytokines with tailored signaling properties to minimize off-target effects. |
| **Scale of Inquiry** | Cellular |
| **Core Question** | How does a cell integrate signals over time to decide its state and fate? |
| **AI-Enabled Problem** | To model the selection dynamics of clonal evolution and the adaptive trajectories of tissue-resident cells, predicting their fate based on signaling history. |
| **Key Data Inputs** | Perturbative single-cell multi-omics; intravital imaging of selection events; Time-series and single-cell omics from defined tissue micro-niches. |
| **Representative AI Models** | Conditional Diffusion Models: By learning the probability distribution of cell fates, these models can run "forwards" to predict outcomes or "backwards" to design the signals required to achieve a desired state. |
| **Therapeutic Opportunity** | Cell State Reprogramming: Computationally designing novel therapeutic protocols (e.g., cytokine cocktails) to guide cells towards desired functional states, such as engineering CAR-T cells that resist exhaustion or generating stable regulatory T cells. |
| **Scale of Inquiry** | Tissue |
| **Core Question** | What are the rules that govern immune cell organization and collective behavior in tissues? |
| **AI-Enabled Problem** | To infer the emergent logic of tissue function by building executable, spatiotemporal models from imaging and spatial omics. |
| **Key Data Inputs** | Spatial transcriptomics & proteomics, multiplex imaging, intravital microscopy |
| **Representative AI Models** | Graph Neural Networks (GNNs): Tissues are fundamentally graphs of interacting cells. GNNs are purpose-built to learn from this structure via "message-passing," a direct analogue of cell-cell communication. |
| **Therapeutic Opportunity** | Cell-Cell Communication Elucidation: Creating in silico models of tissue systems to design therapies that remodel entire tissue ecosystems, disrupting the specific cell-cell interactions that sustain tumor sanctuaries or drive chronic inflammation. |

| Scale of Inquiry | Individual |
|---|---|
| Core Question | How can we predict the trajectory of an individual's immune response and its resolution? |
| AI-Enabled Problem | To build a personalized in silico model by integrating longitudinal, multi-scale patient data. |
| Key Data Inputs | Longitudinal multi-omics, clinical health records, wearable data, imaging |
| Representative AI Models | Multi-modal AI & Generative Models: These architectures are designed to find patterns across fundamentally different data types. Generative models can then forecast realistic future states of the system. |
| Therapeutic Opportunity | Predictive Biomarkers & Clinical Trial Modeling: Forecasting an individual's disease trajectory and likely response to immunotherapy and optimizing therapeutic strategies in silico before treatment to predict efficacy and toxicity. |

**Box 1: Glossary of Key Terms**

Causal Measurement Technologies: Experimental techniques designed to reveal cause-and-effect relationships, rather than just correlations. Examples include CRISPR-based gene editing to assess the functional impact of a specific gene on a cellular process, and perturbation screens where cells are systematically exposed to different stimuli to map their response pathways.

Generative AI Models: A class of artificial intelligence models that can create new data that is similar to the data they were trained on. In biology, these models can be used to design novel proteins with specific functions, generate synthetic biological data for training other models, and predict the outcomes of biological experiments.

Inductive Bias: Assumptions made by a machine learning model to learn the target function and to generalize beyond the training data. In the context of this paper, a biologically informed inductive bias would involve incorporating known biological principles into the architecture of an AI model to guide its learning process.

Predictive Immunology Loop: A proposed framework for accelerating progress in immunology by creating a closed loop between experimental data generation and computational modeling. In this loop, causal experimental data is used to train predictive AI models, which then generate new hypotheses that are tested experimentally, leading to a continuous cycle of learning and discovery.